\documentclass[10pt,english,british,tightenlines,eqsecnum,
floats,aps,amsmath,amssymb,nofootinbib,superscriptaddress,prd,showpacs,showkeys]
              {revtex4}
\usepackage[latin9]{inputenc}
\usepackage{color}
\usepackage{babel}
\usepackage{amsmath}
\usepackage{amssymb}
\usepackage{graphicx}
\usepackage{esint}
\usepackage[unicode=true,pdfusetitle,
 bookmarks=true,bookmarksnumbered=false,bookmarksopen=false,
 breaklinks=false,pdfborder={0 0 1},backref=false,colorlinks=false]
 {hyperref}

\makeatletter
\@ifundefined{textcolor}{}
{%
 \definecolor{BLACK}{gray}{0}
 \definecolor{WHITE}{gray}{1}
 \definecolor{RED}{rgb}{1,0,0}
 \definecolor{GREEN}{rgb}{0,1,0}
 \definecolor{BLUE}{rgb}{0,0,1}
 \definecolor{CYAN}{cmyk}{1,0,0,0}
 \definecolor{MAGENTA}{cmyk}{0,1,0,0}
 \definecolor{YELLOW}{cmyk}{0,0,1,0}
}

\@ifundefined{date}{}{\date{14-Feb-2018}}
\usepackage{babel}

\usepackage{euscript}\usepackage{epsfig}

\usepackage{amsfonts}

\usepackage{fancyhdr}

\makeatother

\begin{document}

\title{Modified Saez-Ballester scalar-tensor theory from $5D$ space-time}

\author{S. M. M. Rasouli}

\email{mrasouli@ubi.pt}

\affiliation{Departamento de F\'{i}sica, Universidade da Beira Interior, Rua Marqu\^{e}s d'Avila
e Bolama, 6200 Covilh\~{a}, Portugal}

\affiliation{Centro de Matem\'{a}tica e Aplica\c{c}\~{o}es (CMA-UBI),
Universidade da Beira Interior, Rua Marqu\^{e}s d'Avila
e Bolama, 6200 Covilh\~{a}, Portugal}

\author{Paulo Vargas Moniz}

\email{pmoniz@ubi.pt}

\affiliation{Departamento de F\'{i}sica, Universidade da Beira Interior, Rua Marqu\^{e}s d'Avila
e Bolama, 6200 Covilh\~{a}, Portugal}

\affiliation{Centro de Matem\'{a}tica e Aplica\c{c}\~{o}es (CMA-UBI),
Universidade da Beira Interior, Rua Marqu\^{e}s d'Avila
e Bolama, 6200 Covilh\~{a}, Portugal}

\begin{abstract}
In this paper, we bring together the five-dimensional
Saez-Ballester~(SB) scalar-tensor theory~\cite{SB85-original} and
the induced-matter-theory~(IMT) setting~\cite{PW92}, to obtain a
modified SB theory (MSBT) in four
dimensions.  Specifically, by using an intrinsic
dimensional reduction procedure into the SB field
equations in five-dimensions, a MSBT
is obtained onto a hypersurface orthogonal to the extra
dimension. This four-dimensional MSBT is shown to bear distinctive
new features in contrast to the usual corresponding SB theory as
well as to IMT and the Modified Brans-Dicke Theory (MBDT)~\cite{RFM14}.
In more detail, besides usual induced matter terms retrieved through the
IMT, the MSBT scalar field is provided with additional physically
distinct (namely, SB induced) terms as well as an
intrinsic self-interacting potential (interpreted as consequence of the IMT
process and the concrete geometry associated to the extra dimension).
 Moreover, our MSBT has four sets of field equations, with two
sets having no analog in the standard SB scalar-tensor theory.
It should be emphasized that the herein appealing solutions can emerge solely from the
geometrical reductional process, from presence also of extra
dimension(s) and not from any ad-hoc matter either in the bulk or on the hypersurface.
 Subsequently, we apply the herein MSBT to cosmology and consider
 an extended spatially flat FLRW geometry in a
five-dimensional vacuum space-time.
After obtaining the exact solutions in the bulk,
we proceed to construct, by
means of the MSBT setting, the corresponding dynamic, on the four-dimensional hypersurface.
More precisely, we obtain the (SB) components of the
induced matter, including the induced scalar potential terms.
We retrieve two different classes of solutions.
Concerning the first class, we show that the MSBT yields a
 barotropic equation of state for the induced perfect fluid.
We then investigate vacuum, dust, radiation, stiff fluid and false vacuum
cosmologies for this scenario and contrast the results with
those obtained in the standard SB theory, IMT and BD theory.
Regarding the second class solutions, we show that the
scale factor behaves similar to a de Sitter (DeS) model.
However, in our MSBT setting, this behavior is assisted by  non-vanishing
induced matter instead, without any a priori cosmological constat.
Moreover, for all these solutions, we show that the extra dimension contracts with the cosmic time.
 \end{abstract}

\medskip



\maketitle

\section{Introduction}
\label{int} \indent

 Higher-dimensional models of the universe have been
widely studied: Kaluza-Klein~(KK) theories~\cite{Kaluza21,Klein26,OW97},
ten-dimensional~and eleven-dimensional supergravity~\cite{Paulo.book1,Paulo.book2,FP12.book} as well as string
theories~\cite{DNP86,GSW.book}.
In particular, Kaluza considered a five-dimensional
space-time, under the following key assumptions~\cite{OW97}, to bring together
electromagnetism with general relativity (GR): (i) there is no ordinary matter in the
higher dimensional space-time, (ii) geometrical quantities similar to those defined in GR were introduced
and (iii) the cylinder condition for the extra coordinate was assumed, which implies that
the derivatives with respect to the extra dimension must vanish.
Following Kaluza's construction, subsequent compactified,
projective~\cite{L82-1,L82-2,S90,S95} and noncompactified~\cite{PW92,OW97,stm99,5Dwesson06,RM16}
 versions
emerged as well-known scenarios
which have been thoroughly investigated, such that, at least, one
of the above mentioned assumptions (i)-(iii) has been modified.
Particularly, by introducing the noncompactified version of the KK theory to
construct the so-called space-time-matter theory (or IMT), it has been
attempted to obtain a firm basis concerning a suggestion by Einstein, who
proposed that it is possible to transpose the ``base wood" of the
matter fields into the `` pure marble" of the geometry in his general
relativistic theory \cite{stm99}.

 Increasingly applications of gravitation
and cosmology versions of KK theory, especially modern KK theories
(namely, the IMT and the brane world scenarios \cite{RS99-1,RS99-2}) have motivated the
investigation of embedding theorems and their corresponding generalized versions.
In this respect, it is important to
establish which embedding
theorem of differential geometry should be considered to
construct a higher dimensional model of gravity.
Historically, within higher-dimensional physical settings, two
embedding theorems have played a pivotal role \cite{DR02-1}.
The first one is the Janet-Cartan theorem \cite{Spivak-book}, in which the embedding space is flat.
The second one, which is particularly more powerful than the first one, is known
as Campbell-Magaard (CM) theorem \cite{C26,M63}, which 
has been
employed to justify, geometrically, the IMT and its modifications \cite{RTZ95,OW97}.
Subsequently, a few extended versions of the CM theorem have been produced.
For instance, in \cite{AL01,ADLR01}, the CM theorem is extended such that
(instead of the Ricci-flatness condition) the embedding
space is the Einstein space sourced with whether one (or more) scalar field(s) or a negative cosmological
constant. Moreover, another extension of this theorem is obtained in \cite{DR02-2}, where it was shown
that the restricted embedding {\it Ricci-flat} space-time can be replaced by a
more general {\it Einstein space-time}. Furthermore, in \cite{SW03}, by
starting from the Gauss-Codazzi equations, another modified proof has been
prepared which can be applicable in the IMT as well as in membrane theory.
We should mention that despite the widespread use in applying the CM
theorem and its generalizations in physical
scenarios, especially in noncompactified KK theories, a few problems
have been mentioned
\cite{A04}
 which has been interpreted as an inadequate ability of the CM theorem concerning the initial-value problem or
non-occurrence of singularities \cite{W05}. As a
defense concerning the mentioned problem, it has been emphasized
that the CM theorem is a local theorem and there is no claim in it about solving
such problems; on the other hand, it has been believed that answering to the problems
such as the singularities as well as causality issues, will be possible
in a complete theory of quantum gravity \cite{W05}.

However, almost all of the mentioned efforts for generalizing the
CM theorem have been done through in a local
\footnote{The first global embedding into $\mathbb{R}^N$ is due to John
Nash \cite{N56} (moreover, as extension appeared in \cite{C70} and \cite{G70}) and a
 recent improvement has been prepared by Gunther \cite{G89}. } setting.
In this respect, Katzourakis proved a global version of the CM theorem
 \cite{K04}, see also \cite{K04-2,K04-3,K04-4}. Subsequently, this embedding theorem has
 been prepared allowing a new global insight to investigate the extra
 force and the particle mass \cite{AB05}.

In this paper, our focus will be on the latter and we will assume that the extra
dimension is noncompact, embracing the motivations of~\cite{stm99,scalarbook,5Dwesson06,pav2006}.
Furthermore, throughout this paper, as a distinctive novel feature, we
will consider, instead of either GR or BD theory,
 the SB scalar-tensor theory~\cite{SB85-original} (as background setting) in a five dimensional
space-time. Then, by using an appropriate reductional
procedure, namely, the IMT algebraic framework, we
will obtain the MSBT induced equations onto a four-dimensional hypersurface.

In the SB theory, in contrast to a BD theory~\cite{BD61} (as well as other ${\rm G}$-varying
theories, see~\cite{Faraoni.book} and references therein), the scalar field does {\em not} play the role of the
(varying) ${\rm G}$. Instead, it is taken as a dimensionless field, not
having to satisfy restrictive constraints determined from observations~\cite{SB85-original}.
Moreover, the strength of the coupling between the gravity and
the scalar field is governed by a dimensionless parameter ${\cal W}$ (see footnote 3).
With such modifications, it has been shown that it
is possible to address the cosmological missing matter problem~\cite{SB85-original}.
In the context of cosmology, by assuming
various line-elements, diverse implications from the SB theory have also been widely employed to obtain
solutions in either four or five dimensions~\cite{P87,SA91,SS03,MSM07,NSR12,Y13,RPR15,RKN12}.
Furthermore, the Noether gauge symmetries of a generalized SB
theory with a scalar potential,
which has been added\footnote{However, in the present paper, we show instead that, by
admitting the SB theory into a five-dimensional
space-time, it is not necessary to add such a potential by hand, but
it can be elegantly retrieved solely from the geometry associated to the extra dimension.}
by hand in~\cite{JAMM12}, has been investigated.

Concerning IMT~\cite{PW92}, it has been shown that matter in a
four-dimensional space-time can be described from a purely
geometric origin. More precisely, the properties of matter, at the macroscopic level, as observed in
the four-dimensional space-time, would be attributed (within a IMT setting) to
the dynamics of a large extra dimension~{\cite{stm99,5Dwesson06,PW92,OW97}.
In cosmology, the IMT has been employed to construct concrete scenarios,
which not only modify the corresponding solutions of GR, but also agree with  recent
cosmological observational data~\cite{stm99,DRJ09,RJ10}.

Applications of the IMT framework, in which the role of GR as a
fundamental underlying theory is replaced by the Brans-Dicke (BD)
scalar-tensor theory~\cite{BD61}, have been formulated in~\cite{ARB07,Ponce1,RFS11,R14,RFM14}.
In the herein paper, we will show that the induced energy-momentum tensor (EMT) in MSBT
 is an extended version of that of IMT, implying significant
differences from gravitational and cosmological perspectives.
In particular, when the scalar field takes constant values we
recover the effective equations of the IMT.
Furthermore, we should note that there are
two extra equations in our herein setting with no analog in the standard SB theory.
As the conventional SB action is a modified version of GR with a canonical scalar
field, our MSBT framework (with the intrinsic scalar potential) can also be employed to study
 inflationary scenarios as well as late time cosmological acceleration.
 These possible scenarios can be considered as the fundamental
 methods to obtain the results. The reason is that
 such possible solutions can be obtained from the geometrical
 implications of the reductional process
 due to extra dimension and not from any ad hoc matter elements.

As a realisation of how MBST can construct an innovative perspective
 in cosmology, we consider the extended
version of the spatially flat Friedmann-Lema\^{\i}tre-Robertson-Walker~(FLRW) line-element in five dimensions. We show
that the geometrically induced matter on the specific hypersurface
 can play the role of ordinary matter.
Moreover, the extra dimension contracts in all of our cosmological solutions.

This paper is organized as follows. In Section~\ref{Set up}, we consider the SB
field equations in five dimensions and then, by employing a dimensional
reduction, extending the IMT procedure, we construct a modified version
of the SB theory on a hypersurface. Moreover, the
modified field equations include new dynamical ingredients, namely, an effective induced
self-interacting SB scalar potential.
In Section~\ref{OT-solution}, by assuming
a five-dimensional vacuum space-time, which is described by the
FLRW line-element, we obtain exact cosmological
solutions for the SB field equations. By employing this MSBT-IMT framework, we will
obtain in section~\ref{OT-reduced} the induced quantities onto the four-dimensional hypersurface. Then, we
study the reduced four-dimensional cosmological solutions.
Subsequently,
we will show that there are two different classes of solutions.
For the first class, we will explain that the resulted solutions can describe
different states of matter fluids in the universe.
Whereas the second class can describe a particular kind of matter
 which leads us to a DeS-like behavior for the scale factor.
Finally, we present our conclusions in Section~\ref{conclusion}.

\section{Dimensional Reduction of Five-dimensional Saez-Ballester Theory}
\label{Set up}
\indent
In this section, by employing a specific dimensional reduction
procedure for a five-dimensional SB theory,
we obtain the effective field equations on a four-dimensional space-time.

The action for the five-dimensional SB theory, in analogy with the
corresponding four-dimensional case~\cite{SB85-original}, can be written as
\begin{equation}\label{SB-5action}
{\cal S}^{^{(5)}}=\int d^{5}x \sqrt{\Bigl|{}{\cal
G}^{^{(5)}}\Bigr|} \,\left[R^{^{(5)}}-{\cal W}\phi^n\,{\cal
G}^{ab}\,(\nabla_a\phi)(\nabla_b\phi)+\chi\,
L\!^{^{(5)}}_{_{\rm matt}}\right],
\end{equation}
where $\phi$ is the dimensionless scalar field and ${\cal W}$ and $n$ are
dimensionless parameters\footnote{We should
notice, in our herein work, the SB action is written in Planck units and that is the reason why
the Einstein-Hilbert term does not have any coefficient and the coupling to
matter is just $\chi=8\pi$. When standard units are used, the scalar field could only
be dimensionless if ${\cal W}$ has dimensions of energy square.}; the Latin indices run from zero
to four; ${\cal G}^{^{(5)}}$ and $R^{^{(5)}}$ are the determinant
and curvature scalar associated with the
five-dimensional metric ${\cal G}_{{ab}}$, respectively.

Here, we should emphasize two points: (i) In contrast to most of the
scalar-tensor theories such as the BD theory~\cite{BD61}, the scalar field does not have
 dimensions of inverse of the (five-dimensional) gravitational
 constant, but instead, it is dimensionless~\cite{SB85-original}.
(ii) In the original setting of IMT~\cite{PW92,OW97}, it was assumed that
there is no ordinary matter in the bulk.
However, for the sake of formal generality, we consider a non-vanishing Lagrangian $L^{^{(5)}}_{_{\rm matt}}$
(which is minimally coupled to the scalar field) describing
 ordinary matter\footnote{In this paper, the ``vacuum'' space-time is used for a situation
where there is~not any type of ordinary matter.}
in the five-dimensional space-time, for the sole reason
of writing broaden applicable expression.

Independently varying action (\ref{SB-5action}) with respect to the
metric and the scalar field gives the equations
\begin{equation}\label{(D+1)-equation-1}
G^{^{(5)}}_{ab}={\cal W}\phi^{n}\left[(\nabla_a\phi)(\nabla_b\phi)
-\frac{1}{2}{\cal G}_{ab}(\nabla^c\phi)(\nabla_c\phi)\right]+\chi\,T^{^{(5)}}_{ab}
\end{equation}
and
\begin{equation}\label{(D+1)-equation-2}
2\phi^n\nabla^2\phi
+n\phi^{n-1}(\nabla_a\phi)(\nabla^a\phi)=0,
\end{equation}
where $\nabla^2\equiv\nabla_a\nabla^a$;
$T^{^{(5)}}_{ab}$ and $G^{^{(5)}}_{ab}$ stand for
the EMT [of the ordinary matter fields] and the Einstein tensor
 in five-dimensional space-time, respectively. From
equation~(\ref{(D+1)-equation-1}), we obtain
\begin{equation}\label{(D+1)-equation-3}
R^{^{(5)}}={\cal W}\phi^{n}(\nabla_a\phi)(\nabla^a\phi)-\frac{2}{3}\chi T^{^{(5)}},
\end{equation}
where $T^{^{(5)}}={\cal G}^{ab}T^{^{(5)}}_{ab}$.

In what follows, by employing the specific reduction procedure, the five-dimensional field
equations will be related to the corresponding ones on the four-dimensional
hypersurface. We will show that effective matter and the induced scalar
potential appearing in the four-dimensional field equations can emerge solely
from the geometry of the fifth dimension.
 Let us be more precise. The effective field equations associated to
 the four-dimensional hypersurface will be derived by
employing the five-dimensional SB field equations~(\ref{(D+1)-equation-1}),
(\ref{(D+1)-equation-2}), and the five-dimensional well-known metric~\cite{PW92,OW97}
\begin{equation}\label{global-metric}
dS^{2}={\cal G}_{ab}(x^c)dx^{a}dx^{b}=
g_{\mu\nu}(x^\alpha,l)dx^{\mu}dx^{\nu}+
\epsilon\psi^2\left(x^\alpha,l\right)dl^{2},
\end{equation}
where $l$ denotes a non-compact coordinate along the extra dimension,
 the Greek indices run from zero to $3$, $\epsilon=\pm1$ permits us to take the extra dimension as
either time-like\footnote{The idea of two-time physics has been highly motivated by the
works by Itzhak Bars and his collaborators~\cite{BDA98,B00-1,B00-2,B08}.
Indeed, the two-time idea in physical theories could provide some insight into
one-time dynamics associated to the higher-dimensional frameworks. The second time has also
been studied in GR when the quality of the universal
parametric ``historical time" has been investigated~\cite{HP73,BH95}.
It has been believed that~\cite{KW04} assigning time-like signature to
the extra coordinate in the KK theories may receive the following constructive criticisms:
(i) assuming the cylinder condition, by integrating over the extra
coordinate in the $5D$ action, the resulted action takes the opposite sign with respect to that of Einstein, which is not correct.
(ii) again, assuming a cylinder condition may yield tachyons. (iii) closed time curves
might emerge.
In our herein MSBT setting, similar to the MBDT and the IMT frameworks, as $\epsilon$ is chosen to
take $\pm1$, assuming the extra coordinate to be time-like is less restricted.
In IMT and membrane theory, two-time physics has also
been the main motivation of the investigation, see, e.g.,~\cite{stm99,BW96,BW96-2,W02}.
 In what follows, let us mention some of
their corresponding consequences. It has been shown that: as the
second time-like coordinate is related to the (inertial) rest
mass of a particle, there is no problem with the criticism (iii)~\cite{W02}; the waves in the
extra dimension, which oscillate around the hypersurface, can be
described by assuming general two-time metric as given in~\cite{W02}.
Therefore, such models can be employed to describe multiply-imaged particles.
As the MBDT~\cite{RFM14} as well as our herein MSBT frameworks are the
generalized versions of the IMT setting, we suggest
that investigating the two-time physics, even by assuming the
same metrics, can yield more generalized consequences.} or space-like and $\psi$ may depend on all the coordinates.
It is usually supposed that the five-dimensional space-time is
foliated by a family of four-dimensional hypersurfaces, $\Sigma$.
For instance, we can set $l=l_{0}={\rm constant}$ to get $\Sigma_0$, whose intrinsic
line-element is orthogonal to the
five-dimensional unit vector
\begin{equation}\label{unitvector}
n^a=\frac{\delta^a_{_4}}{\psi} \qquad {\rm where} \qquad
n_an^a=\epsilon,
\end{equation}
along the extra dimension~\cite{Ponce1,RFM14}.
Consequently, we have the induced metric $g_{\mu\nu}$ on the hypersurface
$\Sigma_{0}$ with the form
\begin{equation}\label{brane-metric}
ds^{2}={\cal G}_{\mu\nu}(x^{\alpha},
l_{0})dx^{\mu}dx^{\nu}\equiv g_{\mu\nu}dx^{\mu}dx^{\nu}.
\end{equation}
Now, in order to get the four-dimensional part of
the corresponding five-dimensional quantity,
we can set $a\rightarrow\mu$ and $b\rightarrow\nu$ in equation~(\ref{(D+1)-equation-1}). Consequently, we obtain
\begin{eqnarray}\label{d+1-Einstein}
G_{\mu\nu}^{^{(5)}}\!\!\!&=&{\cal W}\phi^{n}
\left[({\cal D}_\mu\phi)({\cal D}_\nu\phi)-\frac{1}{2}
g_{\mu\nu}({\cal D}_\alpha\phi)({\cal D}^\alpha\phi)\right]
-\frac{\epsilon{\cal W}\phi^n}{2}\left(\frac{\overset{*}{\phi}}{\psi}\right)^2g_{\mu\nu}+\chi T^{^{(5)}}_{\mu\nu},
\end{eqnarray}
where ${\cal D}_\alpha$ and the notation $\overset{*}A$ stand for the covariant derivative on the
hypersurface (whose computation employs $g_{\mu\nu}$) and the derivative of any quantity
$A$ with respect to the extra coordinate $l$, respectively.
Moreover, ${\cal D}^2\equiv{\cal D}^\alpha{\cal D}_\alpha$ and the following relations have been employed
\begin{eqnarray}\label{rel.1}
(\nabla^c\phi)(\nabla_c\phi)\!\!&=&\!\!({\cal D}^\alpha\phi)({\cal D}_\alpha\phi)+
\epsilon\left(\frac{\overset{*}{\phi}}{\psi}\right)^2,\\
\label{rel.2}
\nabla_\mu\nabla_\nu\phi\!\!&=&\!\!{\cal D}_\mu{\cal D}_\nu\phi+
\frac{\epsilon\overset{*}{\phi}\overset{*}{g}_{\mu\nu}}{2\psi^2},\\
\label{rel.3} \nabla^2\phi\!\!&=&\!\!{\cal D}^2\phi+\frac{({\cal
D}_\alpha\psi)({\cal D}^\alpha\phi)}{\psi}
+\frac{\epsilon}{\psi^2}\left[\overset{**}{\phi}+\overset{*}{\phi}
\left(\frac{g^{\mu\nu}\overset{*}{g}_{\mu\nu}}{2}-\frac{\overset{*}{\psi}}{\psi}\right)\right],\\
\nabla_4\nabla_4\phi\!\!&=&\!\!\epsilon\psi({\cal D}_\alpha\psi)({\cal D}^\alpha\phi)
+\overset{**}{\phi}-\frac{\overset{*}{\psi}}{\psi}\overset{*}{\phi}.
\label{rel.4}
\end{eqnarray}
In order to derive the SB effective field equations on the hypersurface, let us
first construct the left hand side of the field equations, namely the Einstein tensor, on the hypersurface.
Therefore, we have to relate both the Ricci
tensor, $R^{^{(5)}}_{\alpha\beta}$, and the Ricci curvature,
$R^{^{(5)}}$, to their corresponding quantities on the
four-dimensional hypersurface. In this respect, we get
\begin{eqnarray}\label{ricci-tensor-D+1,D}
R^{^{(5)}}_{\alpha\beta}\!\!\!&=&\!\!\!
R^{^{(5)}}_{\alpha\beta}-\frac{{\cal D}_\alpha{\cal
D}_\beta\psi}{\psi}
+\frac{\epsilon}{2\psi^2}\left[\frac{\overset{*}{\psi}\overset{*}{g}_{\alpha\beta}}{\psi}
-\overset{**}{g}_{\alpha\beta}+g^{\lambda\mu}\overset{*}{g}_{\alpha\lambda}\overset{*}{g}_{\beta\mu}
-\frac{1}{2}g^{\mu\nu}\overset{*}{g}_{\mu\nu}\overset{*}{g}_{\alpha\beta}\right],
\\\nonumber
\\
\label{R_DD}
R^{^{(5)}}_{_{44}}\!\!\!&=&\!\!\!-\epsilon\psi{\cal
D}^2\psi-\frac{1}{4}
\overset{*}{g}^{\lambda\beta}\overset{*}{g}_{\lambda\beta}
-\frac{1}{2}g^{\lambda\beta}\overset{**}{g}_{\lambda\beta}
+\frac{1}{2\psi}g^{\lambda\beta}\overset{*}{g}_{\lambda\beta}\overset{*}{\psi}.
\end{eqnarray}
Moreover, by employing equations~(\ref{(D+1)-equation-1}),
(\ref{(D+1)-equation-3}) and
(\ref{R_DD}), we can easily get
\begin{eqnarray}\label{D2say}
\frac{{\cal D}^2\psi}{\psi}=-\frac{\epsilon}{2\psi^2}
\left[g^{\lambda\beta}\overset{**}{g}_{\lambda\beta}
+\frac{1}{2}\overset{*}{g}^{\lambda\beta}\overset{*}{g}_{\lambda\beta}
-\frac{g^{\lambda\beta}\overset{*}{g}_{\lambda\beta}\overset{*}{\psi}}{\psi}\right]
-\frac{\epsilon{\cal W}\phi^n(\overset{*}{\phi})^2}{\psi^2}
+\chi\left[\frac{T^{^{(5)}}}{3}-\frac{\epsilon T^{^{(5)}}_{_{44}}}{\psi^2}\right],
\end{eqnarray}
where we have used $g^{\mu\beta}g^{\lambda\sigma}\overset{*}{g}_{\lambda\beta}
\overset{*}{g}_{\mu\sigma}+\overset{*}g^{\mu\sigma}\overset{*}{g}_{\mu\sigma}=0$.
 Using relations~(\ref{R_DD}) and
(\ref{D2say}), the Ricci scalar in
five-dimensional and four-dimensional space-times are related to each other as
\begin{eqnarray}\label{R(D+1)-R(D)}
R^{^{(5)}}=R^{^{(4)}}-\frac{\epsilon}{4\psi^2}
\left[\left(\overset{*}{g}^{\alpha\beta}\overset{*}{g}_{\alpha\beta}\right)^2
+{\overset{*}{g}^{\alpha\beta}}\overset{*}{g}_{\alpha\beta}\right]
+\frac{2\epsilon{\cal W}\phi^n\overset{*}{\phi}^2}{\psi^2}+2\chi
\left(\frac{\epsilon T^{^{(5)}}_{_{44}}}{\psi^2}-\frac{T^{^{(5)}}}{3}\right).
\end{eqnarray}

Subsequently, by employing (\ref{ricci-tensor-D+1,D})-(\ref{R(D+1)-R(D)}), the
effective field equations (onto the $4D$ hypersurface) associated to
our MSBT setting can be constructed.
In the following three steps, by presenting appropriate
interpretations, we would outline the MSBT field equations.

Firstly, the SB equations on the hypersurface can be constructed by using
equations~(\ref{d+1-Einstein}), (\ref{ricci-tensor-D+1,D})
and (\ref{R(D+1)-R(D)}), as
\begin{eqnarray}\label{BD-Eq-DD}
G_{\mu\nu}^{^{(4)}}\!\!&=&\!\!\chi\left(E_{\mu\nu}+T_{\mu\nu}^{^{[\rm SB]}}\right)+
{\cal W}\phi^n\left[({\cal D}_\mu\phi)({\cal D}_\nu\phi)-
\frac{1}{2}g_{\mu\nu}({\cal D}_\alpha\phi)({\cal
D}^\alpha\phi)\right] -\frac{1}{2}g_{\mu\nu}V(\phi) \cr
 & \equiv &
\chi T_{\mu\nu}^{^{(4)[{\rm eff}]}} +
{\cal W}\phi^n\left[({\cal D}_\mu\phi)({\cal D}_\nu\phi)-
\frac{1}{2}g_{\mu\nu}({\cal D}_\alpha\phi)({\cal
D}^\alpha\phi)\right]-\frac{1}{2}g_{\mu\nu}V(\phi).
\end{eqnarray}

This consequently conveys the standard SB equations on the
four-dimensional hypersureface which comprise an
induced scalar potential, see equation~(\ref{v-def}).
Moreover, in what follows, we should make clear a few points:
\begin{itemize}
\item The effects of the five-dimensional ordinary EMT on
    the hypersurface are denoted by $E_{\mu\nu}$, which is given by
\begin{eqnarray}\label{S}
E_{\mu\nu}\equiv T_{\mu\nu}^{^{(5)}}-
g_{\mu\nu}\left[\frac{T^{^{(5)}}}{3}-
\frac{\epsilon\, T_{_{44}}^{^{(5)}}}{\psi^2}\right].
\end{eqnarray}
 Obviously, if, at the beginning, we assume that there is no ordinary matter fields in the
  five-dimensional space-time [i.e., if we started by assuming $L_{_{\rm matt}}^{(5)}=0$ in action (2.1)],
 then $E_{\mu\nu}$ vanishes.

\item
The quantity $T_{\mu\nu}^{^{[\rm SB]}}$ is an effective EMT
for our MSBT setting, which comprises three components as
\begin{eqnarray}\label{matt.def}
\chi T_{\mu\nu}^{^{[\rm SB]}}= T_{\mu\nu}^{^{[\rm IMT]}}+T_{\mu\nu}^{^{[\rm \phi]}}
+\frac{1}{2}g_{\mu\nu}V(\phi),
\end{eqnarray}
where
\begin{eqnarray}\label{IMTmatt.def}
T_{\mu\nu}^{^{[\rm IMT]}}&\!\!\!\equiv &\!\!
\frac{{\cal D}_\mu{\cal D}_\nu\psi}{\psi}
-\frac{\epsilon}{2\psi^{2}}\left(\frac{\overset{*}{\psi}\overset{*}{g}_{\mu\nu}}{\psi}-\overset{**}{g}_{\mu\nu}
+g^{\lambda\alpha}\overset{*}{g}_{\mu\lambda}\overset{*}{g}_{\nu\alpha}-\frac{1}{2}
g^{\alpha\beta}\overset{*}{g}_{\alpha\beta}\overset{*}{g}_{\mu\nu}\right)\cr
 \!\!\!&&\!\!-\frac{\epsilon g_{\mu\nu}}{8\psi^2}
\left[\overset{*}{g}^{\alpha\beta}\overset{*}{g}_{\alpha\beta}
+\left(g^{\alpha\beta}\overset{*}{g}_{\alpha\beta}\right)^{2}\right],\\\nonumber
\\
\label{T-phi} T_{\mu\nu}^{^{[\rm
\phi]}}&\!\!\!\equiv &\!\!
\left[\frac{\epsilon{\cal W}}{2}\phi^n\left(\frac{\overset{*}{\phi}}{\psi}\right)^2\right]g_{\mu\nu}.
\end{eqnarray}
The first contribution in the effective EMT, i.e. $T_{\mu\nu}^{^{[\rm
IMT]}}$, emerges from the fifth part of the metric (\ref{global-metric}) which is a consequence from the
geometry of the extra dimension;
while the second
contribution $T_{\mu\nu}^{^{[\rm \phi]}}$ depends on the scalar
field $\phi$ and its derivative with respect to the fifth coordinate.

\end{itemize}

Secondly, we should derive the counterpart of
equation~(\ref{(D+1)-equation-2}) onto the four-dimensional space-time. By substituting relations~(\ref{rel.1})
and~(\ref{rel.3}) into equation~(\ref{(D+1)-equation-2}), we obtain
\begin{eqnarray}\label{D2-phi}
2\phi^n{\cal D}^2\phi+n\phi^{n-1}({\cal D}_\alpha\phi)({\cal D}^\alpha\phi)
-\frac{1}{\cal W}\frac{dV(\phi)}{d\phi}=0,
\end{eqnarray}
where
\begin{eqnarray}\label{v-def}
 \frac{dV(\phi)}{d\phi}\equiv-\frac{2{\cal W}\phi^n}{\psi^2}
\Bigg\{\psi({\cal D}_\alpha)({\cal D}^\alpha\phi)
+\frac{n\epsilon}{2}\Big(\frac{\overset{*}{\phi}^2}{\phi}\Big)
+\epsilon\Big[\overset{**}{\phi}+\overset{*}{\phi}
\Big(\frac{g^{\mu\nu}\overset{*}{g}_{\mu\nu}}{2}-\frac{\overset{*}{\psi}}{\psi}\Big)\Big]\Bigg\}.
\end{eqnarray}
It is important to note that, in our herein approach, the
dimensional reduction procedure leads us to a differential equation to get the induced scalar potential.
We should emphasize that, in most of the conventional scalar-tensor theories, such a
potential has instead been added by hand to the action, see, e.g.,~\cite{Faraoni.book,JAMM12}.

Finally, we would introduce the corresponding conservation equation
as that obtained in the IMT. In this respect, by
setting $a\rightarrow\alpha$ and $b\rightarrow D$ in equation~(\ref{(D+1)-equation-1}), we obtain
\begin{eqnarray}\label{G-D,alpha}
G_{\alpha 4}^{^{(5)}}=R_{\alpha 4}^{^{(5)}}=\chi T^{^{(5)}}_{\alpha 4}+{\cal W}\phi^n\overset{*}{\phi}({\cal
D}_\alpha\phi),
\end{eqnarray}
where, to get the first equality, we have used the metric~(\ref{global-metric}).
Again, using (\ref{global-metric}) gives~\cite{PW92}
\begin{equation}\label{GDmo-2}
G^{^{(5)}}_{\alpha 4}= \psi P^{\beta}{}_{\alpha;\beta},
\end{equation}
where $P_{\alpha\beta}$ was defined as
\begin{equation}\label{P-mono}
P_{\alpha\beta}\equiv\frac{1}{2
\psi}\left(\overset{*}{g}_{\alpha\beta}
-g_{\alpha\beta}g^{\mu\nu}\overset{*}{g}_{\mu\nu}\right).
\end{equation}
Consequently, using equations~(\ref{G-D,alpha}) and (\ref{GDmo-2}), we
get a dynamical equation for $P_{\alpha\beta}$ as
\begin{eqnarray}\label{P-Dynamic}
\psi P^{\beta}{}_{\alpha;\beta}&=&\!\!
\chi T^{^{(5)}}_{\alpha 4}+{\cal W}\phi^n\overset{*}{\phi}({\cal
D}_\alpha\phi)\,.
\end{eqnarray}

As we close this section, let us further elucidate a few
concepts of the herein retrieved four-dimensional MSBT.
\begin{itemize}
\item
 The field equations associated to the five-dimensional
 SB theory, i.e., (\ref{(D+1)-equation-1}) and (\ref{(D+1)-equation-2}),
with the metric (\ref{global-metric}), yield
four sets of the effective field equations~(\ref{D2say}), (\ref{BD-Eq-DD}), (\ref{D2-phi})
and (\ref{P-Dynamic}) on the hypersurface.
Indeed, from the point of view of an observer in the
four-dimensional space-time, who is not aware of the
existence of the fifth dimension, equations~(\ref{BD-Eq-DD})
and~(\ref{D2-phi}) could be interpreted as the field equations associated to
conventional SB theory in four dimensions.
Such a description can be considered as a consequence of the generalized versions of the CM
 theorem~\cite{AL01,ADLR01,DR02-2}.
More precisely, the field equations (\ref{BD-Eq-DD}) and
(\ref{D2-phi}) could be obtained from the
action
\begin{equation}\label{induced-action}
 {\cal S}^{^{(4)}}=\int d^{^{\,4}}\!x \sqrt{-g}\,\left[R^{^{(4)}}-{\cal W}\phi^n\, g^{\alpha\beta}\,({\cal
D}_\alpha\phi)({\cal D}_\beta\phi)-V(\phi)+\chi\,
L\!^{^{(4)}}_{_{\rm matt}}\right],
\end{equation}
 where
\begin{equation}\label{induced-source}
\sqrt{-g}\left(E_{\alpha\beta}+T^{^{[\rm
SB]}}_{\alpha\beta}\right)\equiv 2\delta\left( \sqrt{-g}\,
L\!^{^{(4)}}_{_{\rm matt}}\right)/\delta g^{\alpha\beta}.
 \end{equation}
Moreover, we should note that equation~(\ref{D2say}) has no analog in the
conventional SB theory, and the set of equations~(\ref{P-Dynamic}) is an
extended version of the conservation law obtained in the IMT.
\item
 It is straightforward to show that the effective EMT is covariantly conserved,
i.e., ${\cal D}_\beta T^{^{[\rm SB]}}_\alpha{}^{\beta}=0$.

\item
 In the particular case when $\phi={\rm constant}$
 and $T^{^{(5)}}_{\alpha 4}=0$, equation~(\ref{P-Dynamic}) reduces to
$P^{^\beta}{}_{\alpha;\beta}=0$. Moreover, if $l$ is assumed to be a
cyclic coordinate, then (\ref{P-Dynamic}) reduces to an identity.
\end{itemize}

\section{exact solutions of SB cosmology in five-Dimensional vacuum space-time}
\label{OT-solution}

\indent
In this section, we shall work in a five-dimensional space-time and obtain exact cosmological solutions.
 Subsequently, in the next section, by using the MSBT setting, we will derive the effective
 EMT onto a four-dimensional hypersurface and will analyse the corresponding solutions.

Let us start by assuming a five-dimensional vacuum
space-time described by an extended version of
the~spatially flat FLRW line-element as
\begin{equation}\label{DO-metric}
dS^{2}=-dt^{2}+a^{2}(t)\left[dr^2+r^2\left(d\theta^2+sin^2\theta d\varphi^2\right)\right]
+\epsilon \psi^2(t)dl^{2},
\end{equation}
where $t$ is the cosmic time and $(r,\theta,\varphi)$ are the
spherical coordinates. Due to the space-time
symmetries, we assume that $a$, $\psi$ and $\phi$
depend only on the
comoving time\rlap.\footnote{In this work, let us leave aside
the modified results which can be retrieved by considering the
following cases: (i) taking the scale factors
and the scalar field which depend also on the
spatial coordinates, specially $l$,
 (ii) considering ordinary matter fields in five-dimensional space-time.
 (By performing the calculations without (ii), the proposal ``replacing the
base wood of matter by the pure marble of geometry" is satisfied~\cite{DNP86}.),
(iii) Obtaining the solutions associated to the other values of the spatial curvature constant.}

Therefore, employing the field equation~(\ref{(D+1)-equation-1}) and (\ref{(D+1)-equation-2})
 for the metric~(\ref{DO-metric}), in
vacuum (please see the footnote~4), we obtain
\begin{eqnarray}\label{ohanlon-eq-1}
\frac{\ddot{\phi}}{\phi}+\left[3\left(\frac{\dot{a}}{a}\right)
+\frac{n}{2}\left(\frac{\dot{\phi}}
{\phi}\right)+\frac{\dot{\psi}}{\psi}\right]\frac{\dot{\phi}}{\phi}=0,\\
\label{ohanlon-eq-2}
\frac{\dot{a}}{a}\left(\frac{\dot{a}}{a}+\frac{\dot{\psi}}{\psi}\right)=\frac{{\cal W}}{6}\phi^n\dot{\phi}^2,\\
\label{ohanlon-eq-3}
\frac{2\ddot{a}}{a}+\frac{\dot{a}}{a}
\left(\frac{\dot{a}}{a}+\frac{2\dot{\psi}}{\psi}\right)
+\frac{\ddot{\psi}}{\psi}=-\frac{{\cal W}}{2}\phi^n\dot{\phi}^2,\\
\label{ohanlon-eq-4}
\frac{\ddot{a}}{a}+\left(\frac{\dot{a}}{a}\right)^2
=-\frac{{\cal W}}{6}\phi^n\dot{\phi}^2.
\end{eqnarray}
where ``\,\,${\bf\dot{}}$\,\,'' denotes the derivative
with respect to the cosmic time.

Among the coupled non-linear field
equations~(\ref{ohanlon-eq-1})-(\ref{ohanlon-eq-4}), only three of them are independent
and we have three unknown quantities $a(t)$, $\phi(t)$ and $\psi(t)$.
From equation (\ref{ohanlon-eq-1}), we get a constant of motion as
\begin{eqnarray}\label{new1}
a^3\dot{\phi}\phi^{\frac{n}{2}}\psi=c_1,
\end{eqnarray}
where $c_1$ is a constant of integration.
Moreover, using (\ref{ohanlon-eq-2}) and (\ref{ohanlon-eq-3}), we get
\begin{eqnarray}\label{31}
\frac{{\cal W}}{6}\phi^n\dot{\phi}^2-\frac{\dot{a}}{a}\left[\frac{\dot{a}}{a}
-2\left(\frac{\dot{\psi}}{\psi}\right)\right]+\frac{\ddot{\psi}}{\psi}=0.
\end{eqnarray}
Now, by employing (\ref{ohanlon-eq-4}) and (\ref{31}) to remove $\frac{{\cal W}}{6}\phi^n\dot{\phi}^2$, we obtain
\begin{eqnarray}\label{32}
3\left(\frac{\dot{a}}{a}\right)\left(\frac{\dot{\psi}}{\psi}\right)+\frac{\ddot{\psi}}{\psi}=0.
\end{eqnarray}
From the above equation, we conclude that $a^3\dot{\psi}$ is another constant of motion. Namely, we have
\begin{eqnarray}\label{34}
a^3\dot{\psi}=c_2,
\end{eqnarray}
where $c_2$ is another constant of integration.
Combining (\ref{new1}) and (\ref{34}), we obtain
\begin{equation}\label{new7}
\psi=\left \{
 \begin{array}{c}
  \psi_0 {\rm exp}\left(\frac{2\beta}{n+2}\phi^{\frac{n+2}{2}}\right)
 \hspace{12mm} {\rm for}\hspace{5mm} n\neq-2,\\\\
 \psi_0\phi^\beta
  \hspace{33mm} {\rm for}\hspace{5mm} n=-2,
 \end{array}\right.
\end{equation}
where $\beta\equiv\frac{c_2}{c_1}$ and $\psi_0$ is a constant of integration.
By employing (\ref{new7}) in (\ref{ohanlon-eq-2}), we get
\begin{equation}\label{new17}
a=\left \{
 \begin{array}{c}
  a_0 {\rm exp}\left(\frac{2\gamma}{n+2}\phi^{\frac{n+2}{2}}\right)
 \hspace{12mm} {\rm for}\hspace{5mm} n\neq-2,\\\\
 a_0\phi^\gamma
  \hspace{33mm} {\rm for}\hspace{5mm} n=-2,
 \end{array}\right.
\end{equation}
where $a_0$ is an integration constant and $\gamma$ was defined as
\begin{eqnarray}\label{new17-2}
\gamma\equiv\frac{1}{2}\left(-\beta\pm\sqrt{\beta^2+\frac{2{\cal W}}{3}}\right).
\end{eqnarray}
By substituting $\psi$ and $a$ from (\ref{new7}) and (\ref{new17}) into (\ref{new1}), we get
\begin{equation}\label{new18-2}
\left \{
 \begin{array}{c}
  \dot{\phi}\phi^{\frac{n}{2}}{\rm exp}\left[\frac{2(3\gamma+\beta)}{n+2}\phi^{\frac{n+2}{2}}\right]=\frac{c_1}{a_0^3\psi_0}
 \hspace{12mm} {\rm for}\hspace{5mm} n\neq-2,\\\\
   \dot{\phi}\phi^{(3\gamma+\beta)-1}=\frac{c_1}{a_0^3\psi_0}
  \hspace{30mm} {\rm for}\hspace{5mm} n=-2.
 \end{array}\right.
\end{equation}
Each of the above differential equations can give two different
solutions whether the quantity $3\gamma+\beta$ vanishes or not.
Integrating, disregarding the values of $n$, we get two different
solutions as presented in the following subsections.
\subsection{Case~I$:3\gamma+\beta\neq0$}
For this case, from (\ref{new18-2}), we obtain
\begin{equation}\label{new20-22}
\left \{
 \begin{array}{c}
 \phi^{\frac{n+2}{2}}={\rm ln}\left[\frac{c_1(3\gamma+\beta)}{a_0^3\psi_0}(t-t_i)\right]^{\frac{n+2}{2(3\gamma+\beta)}}
 \hspace{12mm} {\rm for}\hspace{5mm} n\neq-2,\\\\
   \phi(t)=\left[\frac{c_1(3\gamma+\beta)}{a_0^3\psi_0}(t-t_i)\right]^{\frac{1}{3\gamma+\beta}}
  \hspace{20mm} {\rm for}\hspace{5mm} n=-2,
   \end{array}\right.
\end{equation}
where $t_i$ is an integration constant. Consequently, by substituting (\ref{new20-22})
into relations (\ref{new7}) and (\ref{new17}), we can rewrite
 $a$ and $\psi$ in terms of the cosmic time as
\begin{eqnarray}\label{new29}
 a(t)&=&a_0\left[\frac{c_1(3\gamma+\beta)}{a_0^3\psi_0}(t-t_i)\right]^{\frac{\gamma}{3\gamma+\beta}},\\\nonumber
 \\\label{new29-2}
   \psi(t)&=&\psi_0\left[\frac{c_1(3\gamma+\beta)}{a_0^3\psi_0}(t-t_i)\right]^{\frac{\beta}{3\gamma+\beta}}.
 \end{eqnarray}
 We notice that the above relations do not depend on $n$.

\subsection{Case~II:\,$3\gamma+\beta=0$}
In this case, using (\ref{new17-2}), we can express $\gamma$ and ${\cal W}$ versus $\beta$ as
\begin{eqnarray}\label{Gamma-W}
\gamma=-\frac{\beta}{3}, \hspace{10mm} {\cal W}=-\frac{4}{3}\beta^2.
\end{eqnarray}

 Using (\ref{new18-2}), we get
\begin{equation}\label{new54}
\left \{
 \begin{array}{c}
 \phi^{\frac{n+2}{2}}=\frac{c_1(n+2)}{2a_0^3\psi_0}(t-t_i)
 \hspace{12mm} {\rm for}\hspace{5mm} n\neq-2,\\\\
   \phi(t)={\rm exp}\left[\frac{c_1}{a_0^3\psi_0}(t-t_i)\right]
  \hspace{7mm} {\rm for}\hspace{5mm} n=-2.
 \end{array}\right.
\end{equation}
Therefore, from relations (\ref{new7}), (\ref{new17}) and (\ref{new54}),
we obtain $a$ and $\psi$, in terms of cosmic time, as
\begin{eqnarray}\label{new59}
 a(t)\!\!&=&\!\!a_0\,{\rm exp}\left[-\frac{c_1\beta}{3a_0^3\psi_0}(t-t_i)\right] ,\\
 \label{new59-2}
 \psi(t)\!\!&=&\!\!\psi_0\,{\rm exp}\left[\frac{c_1\beta}{a_0^3\psi_0}(t-t_i)\right].
\end{eqnarray}
Again, we should note that, when the scale factors $a$ and $\psi$ are expressed in terms of the
cosmic time, likewise Case~I, they are independent of $n$.

\section{Reduced Saez-Ballester Cosmology in Four Dimensions}
\label{OT-reduced}
\indent

In this section, by employing the MSBT setting outlined in section \ref{Set up},
we investigate exact cosmological solutions generated on
the four-dimensional hypersurface, which will be compared
with corresponding ones obtained from GR, IMT, standard SB theory and the MBDT.

It is straightforward to show that the non-vanishing
components of the effective EMT (\ref{matt.def})
associated to the line-element~(\ref{DO-metric}),
on the four-dimensional hypersurface $\Sigma_0$, are given by
\begin{eqnarray}\label{R16}
\chi T^{0[{\rm SB}]}_{\,\,\,0}\!\!\!&=&\!\!\!
-\frac{\ddot{\psi}}{\psi}+\frac{1}{2}V(\phi),\\\nonumber
\\
\label{R17}
\chi T^{i[{\rm SB}]}_{\,\,\,i}\!\!\!&=&\!\!\!
-\frac{\dot{a}\dot{\psi}}{a\psi}+\frac{1}{2}V(\phi),
\end{eqnarray}
where $i=1,2,3$ (with no sum) and the induced scalar potential is obtained by employing (\ref{v-def}).
As $\phi$ and $\psi$ only depend on the cosmic time, we get
\begin{equation}\label{new30}
\frac{dV}{d\phi}{\Biggr|}_{_{\Sigma_{o}}}\!\!\!\!\!=2{\cal W}\phi^n\dot{\phi}\left(\frac{\dot{\psi}}{\psi}\right).
\end{equation}
On the other hand, using (\ref{new7}), we get
\begin{equation}\label{new31}
\frac{dV}{d\phi}{\Biggr|}_{_{\Sigma_{o}}}=\left \{
 \begin{array}{c}
 2\beta{\cal W}\left(\phi^{\frac{n}{2}}\dot{\phi}\right)^2\phi^{\frac{n}{2}}
 \hspace{12mm} {\rm for}\hspace{5mm} n\neq-2,\\\\
    2\beta{\cal W}\left(\frac{\dot{\phi}}{\phi}\right)^2\phi^{-1}
  \hspace{12mm} {\rm for}\hspace{5mm} n=-2.
 \end{array}\right.
\end{equation}
In order to obtain the quantity in the parentheses of relations (\ref{new31}),
we should use (\ref{new18-2}). However, as
we get different solutions for the cases $3\gamma+\beta\neq0$
and $3\gamma+\beta=0$, in similarity to the previous section,
we prefer to proceed the analysis in two separated subsections as follows.
\subsection{Case~I:\,$3\gamma+\beta\neq0$}
In this case, by employing relations (\ref{new18-2}) into corresponding relations in
(\ref{new31}), after integrating the resulted equations, we obtain for the
induced potential (which we emphasize is not added by hand and solely emerges through the redaction process)
\begin{equation}\label{new32}
V(\phi)=\left \{
 \begin{array}{c}
V_0\,{\rm exp}\left[\frac{-4(3\gamma+\beta)}{n+2}\phi^{\frac{n+2}{2}}\right]
\hspace{12mm} {\rm for}\hspace{5mm} n\neq-2,\\\\
 V_0\,\phi^{-2(3\gamma+\beta)}
\hspace{28mm} {\rm for}\hspace{5mm} n=-2,
 \end{array}\right.
\end{equation}
where $V_0$ is given by
\begin{equation}\label{V0}
V_0\equiv-\frac{{\cal W}c_1^2\beta}{a_0^6\psi_0^2(3\gamma+\beta)}.
 \end{equation}

Finally, by replacing the scalar field from the corresponding relations in (\ref{new20-22}),
the induced scalar potential can be given versus the cosmic time as
\begin{equation}\label{new40}
V(t)=-\frac{{\cal W}\beta}{(3\gamma+\beta)^3}(t-t_i)^{-2},
\end{equation}
which is independent of $n$.

Now, we can calculate the non-vanishing components of the induced matter for this case.
By employing relations~(\ref{new29}), (\ref{new29-2}) and (\ref{new40})
into (\ref{R16}) and (\ref{R17}), it is easy to show that
\begin{eqnarray}\label{new46}
\rho_{_{\rm SB}}\!\!&=&\!\!\frac{\beta\left[{\cal W}-
6\gamma(3\gamma+\beta)\right]}{2\chi(3\gamma+\beta)^3}\,(t-t_i)^{-2},\\
\label{new48}
p_{_{\rm SB}}\!\!&=&\!\!-\frac{\beta\left[{\cal W}+
2\gamma(3\gamma+\beta)\right]}{2\chi(3\gamma+\beta)^3}\,(t-t_i)^{-2},
\end{eqnarray}
where $\rho_{_{\rm SB}}\equiv T^{0[{\rm SB}]}_{\,\,\,0}$
and $p_{_{\rm SB}}=p_i\equiv T^{i[{\rm SB}]}_{\,\,\,i}$ are
the induced energy density and isotropic pressures, respectively.
We see that these expressions, refereing to geometrical induced
effective matter, are independent of $n$ and
yield a effective (induced) barotropic equation of state, associated
with such induced perfect fluid for this case. Namely,
\begin{eqnarray}\label{new49}
p_{_{\rm SB}}=W_{_{\rm SB}}\rho_{_{\rm SB}}, \hspace{10mm} {\rm where} \hspace{10mm}
W_{_{\rm SB}}=-\frac{{\cal W}+2\gamma(3\gamma+\beta)}{{\cal W}-6\gamma(3\gamma+\beta)}.
\end{eqnarray}
By employing relations (\ref{new17-2}), (\ref{new46}) and (\ref{new48}), it is easy to show that
the induced matter (on the four-dimensional hypersurface) is conserved; namely,
$\dot{\rho}_{_{\rm SB}}+3H(\rho_{_{\rm SB}}+p_{_{\rm SB}})=0$.

In the rest of this subsection, we will concentrate on particular cases for the
 equation of state parameter for $3\gamma+\beta\neq0$.
After obtaining the properties of the reduced quantities as well as geometrical induced
matter and scalar potential, we will compare the results with those
resulted from the standard theories and observational data.

\subsubsection{(Effective) dust cosmologies}
For a relativistic pressureless fluid, by setting ${\rm W}_{_{\rm SB}}=0$ in (\ref{new49}) and using (\ref{new17-2}),
we get\footnote{In what follows, for each
 case, the static cosmological solution with $\gamma=0$ is the trivial result.
 We should note that we will not consider this special case in this paper.} $\gamma=-\frac{2\beta}{3}$ and ${\cal W}=-\frac{4\beta^2}{3}$. Consequently, employing
~(\ref{new20-22}), (\ref{new29}),~(\ref{new29-2}), (\ref{new46}), (\ref{new48}) and~(\ref{new32}), we obtain
\begin{eqnarray}\nonumber
 ds^2\!\!&=&\!\!dS^2|_{_{\Sigma_{o}}}=-dt^2+
 a_0^2\left[\frac{-c_1\beta}{a_0^3\psi_0}(t-t_i)\right]^{\frac{4}{3}}
\left[dr^2+r^2\left(d\theta^2+sin^2\theta d\varphi^2\right)\right],\\\nonumber
 \psi(t)\!\!&=&\!\!\psi_0\left[\frac{-c_1\beta}{a_0^3\psi_0}(t-t_i)\right]^{-1},\\\nonumber
 \phi^{\frac{n+2}{2}}\!\!&=&\!\!{\rm ln}\left[\frac{-c_1\beta}{a_0^3\psi_0}(t-t_i)\right]^{-\frac{n+2}{2\beta}},
  \hspace{5mm} V(\phi)=V_0\,{\rm exp}\left[\frac{4\beta}{n+2}\phi^{\frac{n+2}{2}}\right]\hspace{5mm}
  {\rm for}\hspace{5mm} n\neq-2,\\\nonumber
   \phi(t)\!\!&=&\!\!\left[\frac{-c_1\beta}{a_0^3\psi_0}(t-t_i)\right]^{-\frac{1}{\beta}}, \hspace{12mm} V(\phi)=V_0\,\phi^{2\beta}
  \hspace{26mm} {\rm for}\hspace{5mm} n=-2,\\\nonumber
  V(t)\!\!&=&\!\!-\frac{4}{3}(t-t_i)^{-2},\\
  \label{new90}
    \rho_{_{\rm SB}}(t)\!\!&=&\!\!\frac{8}{3\chi}(t-t_i)^{-2}.
   \end{eqnarray}
   where $V_0$ is given by
\begin{equation}\label{V0-dst}
V_0\equiv-\frac{1}{3}\left(\frac{2\beta c_1}{a_0^3\psi_0}\right)^2.
 \end{equation}
   We observe that for a matter-dominated case, our model yields
  two disconnected branches\rlap.\footnote{The pre-big bag cosmology~\cite{LWC00} has been
  inspired by superstring theory. Such a scenario has also been constructed
  by a class of solutions associated to the low-energy string theory, see, e.g.,
  the Nariai solution associated to the dust fluid in the context of the BD theory~\cite{N63,GFR73,Faraoni.book}, such that for one of
  them (branch I), the scale factor contracts
  when $t<t_i$ and for the other (branch II) it expands for $t>t_i$.
  Whilst, for the branch I, the fifth dimension expands and for the branch II, it contracts.
  We should note that the mentioned behaviors directly
  depend on the integration constants $c_1$, $\beta$, $a_0$ and $\psi_0$.
  We have shown that, by assuming $a_0^3\psi_0>0$, the sign
  of the $c_1\beta$ determines which branch is obtained. Namely, $c_1\beta>0$ yields the
  branch I, while $c_1\beta<0$ yields the branch II.}


However, for both of the branches, by assuming $a_0^3\psi_0>0$, the time behavior of the scalar
field depends not only on the sign of the $c_1\beta$ but also on the signs of $c_1$ and $\beta$.
We have shown that for different values of $n$, dependent on the signs of $c_1$ and $\beta$, the
scalar field can whether contracts or expands with the cosmic time.
For instance, in figure~\ref{phi-dust}, we have plotted the behavior of the
scalar field versus cosmic time for $n=\pm2$.
\begin{figure}
\centering{}
\includegraphics[width=2.8in]{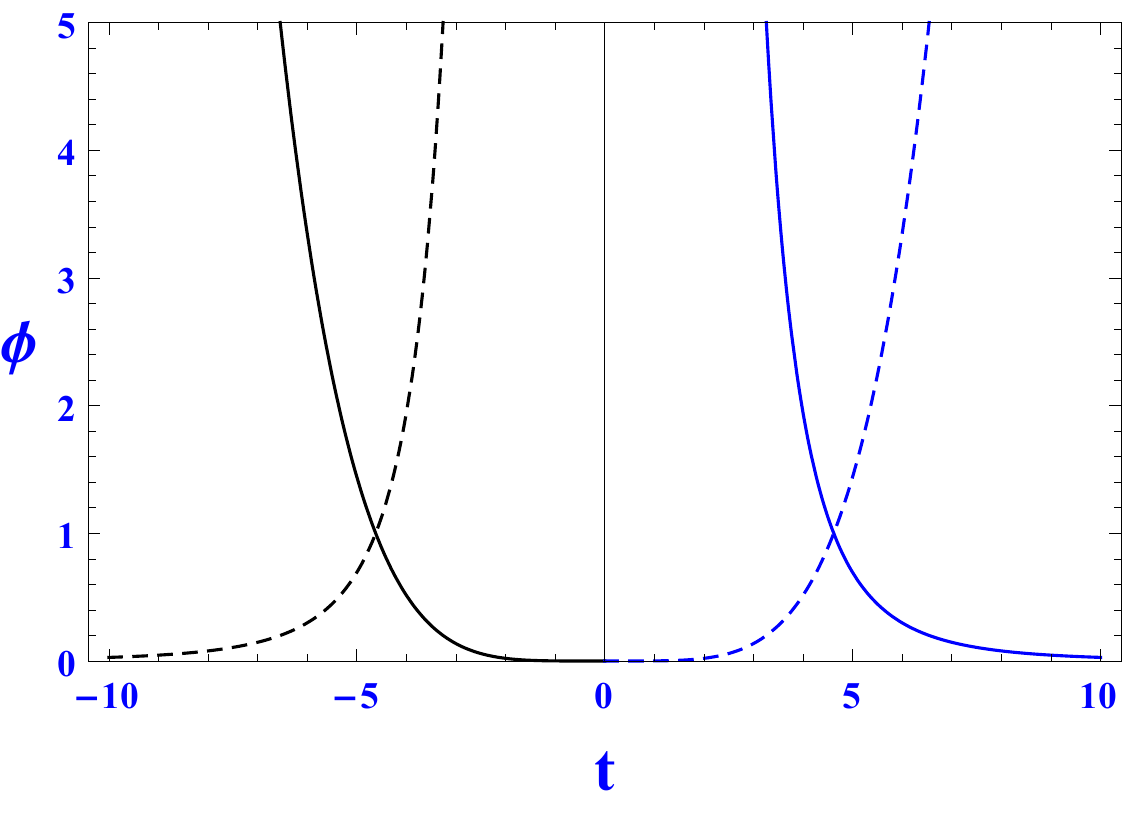}\hspace{4mm}
\includegraphics[width=2.8in]{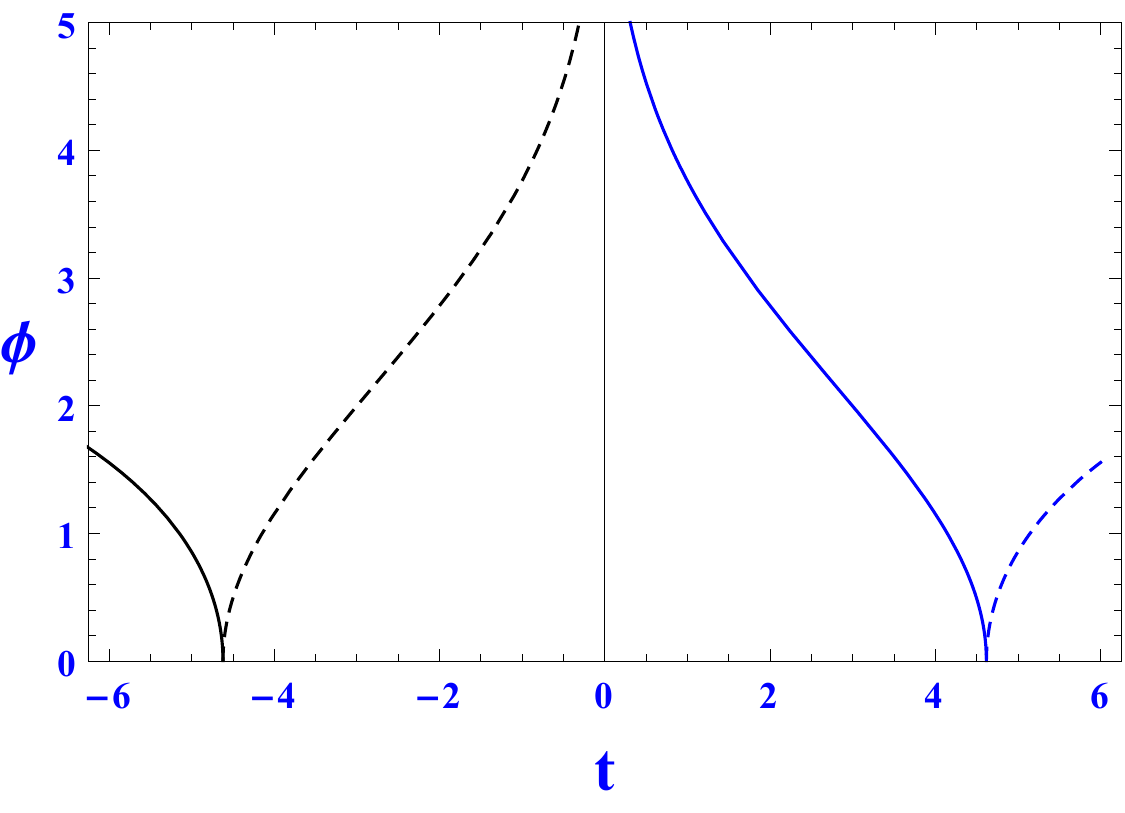}\hspace{4mm}
\caption{{\footnotesize The time behavior of the scalar field
for $n=-2$ (left panel) and $n=2$ (right panel) associated to the dust cosmologies.
We have taken $t_i=0$, $a_0=\psi_0=1$ and $c_1\beta=\pm\sqrt{3}/8$.
Such that the upper and lower
signs are associated to the curves plotted in the
intervals $t<0$ and $t>0$, respectively. Moreover, the dashed
and solid curves correspond to $c_1>0$ and  $c_1<0$, respectively.
.}}
\foreignlanguage{english}{\label{phi-dust}}
\end{figure}

Let us focus on the branch II in which~$t>t_i$. In this case, we have shown that: (i) the scale
factor $a$ is always in a decelerating expansion regime,
which is inapplicable to the present epoch.
(ii) However, the extra dimension and the induced energy density
contract with the cosmic time, which are desirable behaviors
for induced matter theories. (iii) The time behavior of the scalar
field depends not only on the integration constants but also on $n$.
Such that for $n\neq-2$, the
cosmic time must be restricted as $t\geq t_i>0$; namely, the big bang singularity is removed.
Whilst, for $n=-2$, the universe has a big bang singularity.

\subsubsection{(Effective) vacuum cosmologies}

 One of the crucial questions in studying
 cosmological scenarios in the context of scalar-tensor
 theories is whether non-trivial
vacuum solutions
exist or not. For instance, in the context of the BD theory, such solutions are interesting because
 when the time asymptotically goes to zero, for wide ranges of the varying
 BD coupling parameter, the (non-vacuum) FLRW solutions approach the
 vacuum ones~\cite{B93}. Moreover, by studying the cosmological
non-static vacuum solutions (say, for spatially
flat FLRW metric), we can address the validity of the Birkhoff theorem and
Mach's principle in the corresponding context.

In what follows, we would study a particular case in
which the geometrical induced matter vanishes onto
the four-dimensional hypersurface.

 In order to investigate the vacuum solution, from relations~(\ref{new46}) and~(\ref{new48}), we
 find that the only allowed solution
 is obtained by setting $\beta=0$, such that by employing (\ref{new17-2}),
 we get $\gamma=\pm\sqrt{{\cal W}/6}$. Consequently, the
 solutions~(\ref{new20-22}), (\ref{new29}),~(\ref{new29-2}) and ~(\ref{new32}) reduce to
 \begin{eqnarray}\nonumber
 ds^2\!\!&=&\!\!dS^2|_{_{\Sigma_{o}}}=-dt^2+
 \left[\frac{3c_1\gamma}{\psi_0}(t-t_i)\right]^{\frac{2}{3}}\left[dr^2+r^2\left(d\theta^2+sin^2\theta d\varphi^2\right)\right],\\\nonumber
 \psi(t)\!\!&=&\!\!\psi_0={\rm constant},\\\nonumber
 \phi^{\frac{n+2}{2}}\!\!&=&\!\!{\rm ln}\left[\frac{3c_1\gamma}{a_0^3\psi_0}(t-t_i)\right]^{\frac{n+2}{6\gamma}}
 \hspace{12mm} {\rm for}\hspace{5mm} n\neq-2,\\\nonumber
   \phi(t)\!\!&=&\!\!\left[\frac{3c_1\gamma}{a_0^3\psi_0}(t-t_i)\right]^{\frac{1}{3\gamma}}
  \hspace{20mm} {\rm for}\hspace{5mm} n=-2,\\
    \label{new84}
  V(t)\!\!&=&\!\!0,
   \end{eqnarray}
where, for the canonical (scalar field) case, which is obtained by setting $n=0$, we get
\begin{eqnarray}\label{vacuum-canonical}
 \phi={\rm ln}\left[\frac{3c_1\gamma}{a_0^3\psi_0}(t-t_i)\right]^{\frac{1}{3\gamma}},
   \end{eqnarray}
while the other quantities do not change.

Comparing the relations associated to the scalar
fields in (\ref{new90}) and (\ref{new84}), we find that
 by setting\footnote{We should note that the $\beta$ in this case is
 not exactly equal to that in the dust case.} $\gamma\mapsto-\tilde{\beta}/3$
 in the latter, we get the same functions of the scalar field in the former.
 Consequently, by choosing suitable initial values,
  figures~\ref{phi-dust} can also show the behaviors of the
 scalar fields associated to the vacuum solution.

Again, we would focus on the time range where $t>t_i$. Similar to the previous case, for $n\neq-2$,
the relation associated to the scalar field indicates that
the allowed values for the cosmic time is $t\geq t_i>0$. Therefore, for this case, the
relation for the scale factor shows that the universe cannot be started with vanishing size.
Namely, for this case, the big bang singularity is removed. Whereas, for the
case $n=-2$, the quantity $t-t_i$ can
vanish and therefore the universe has a big bang singularity.
In figure~\ref{vac}, we have plotted the behaviors of the
scalar field versus cosmic time for differed vales of $n$ and of the integration constants.
 \begin{figure}
\centering{}\includegraphics[width=2.8in]{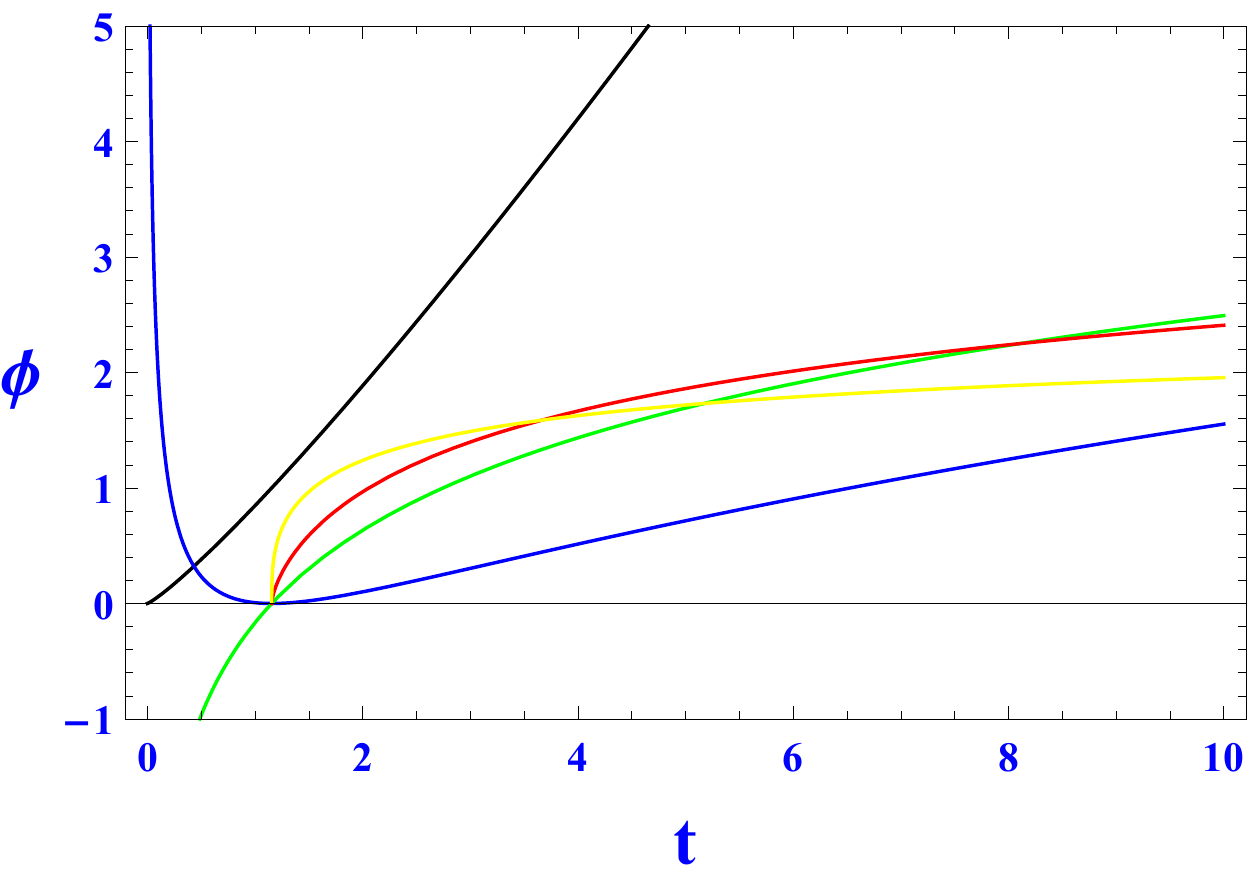}\hspace{4mm}
\includegraphics[width=2.8in]{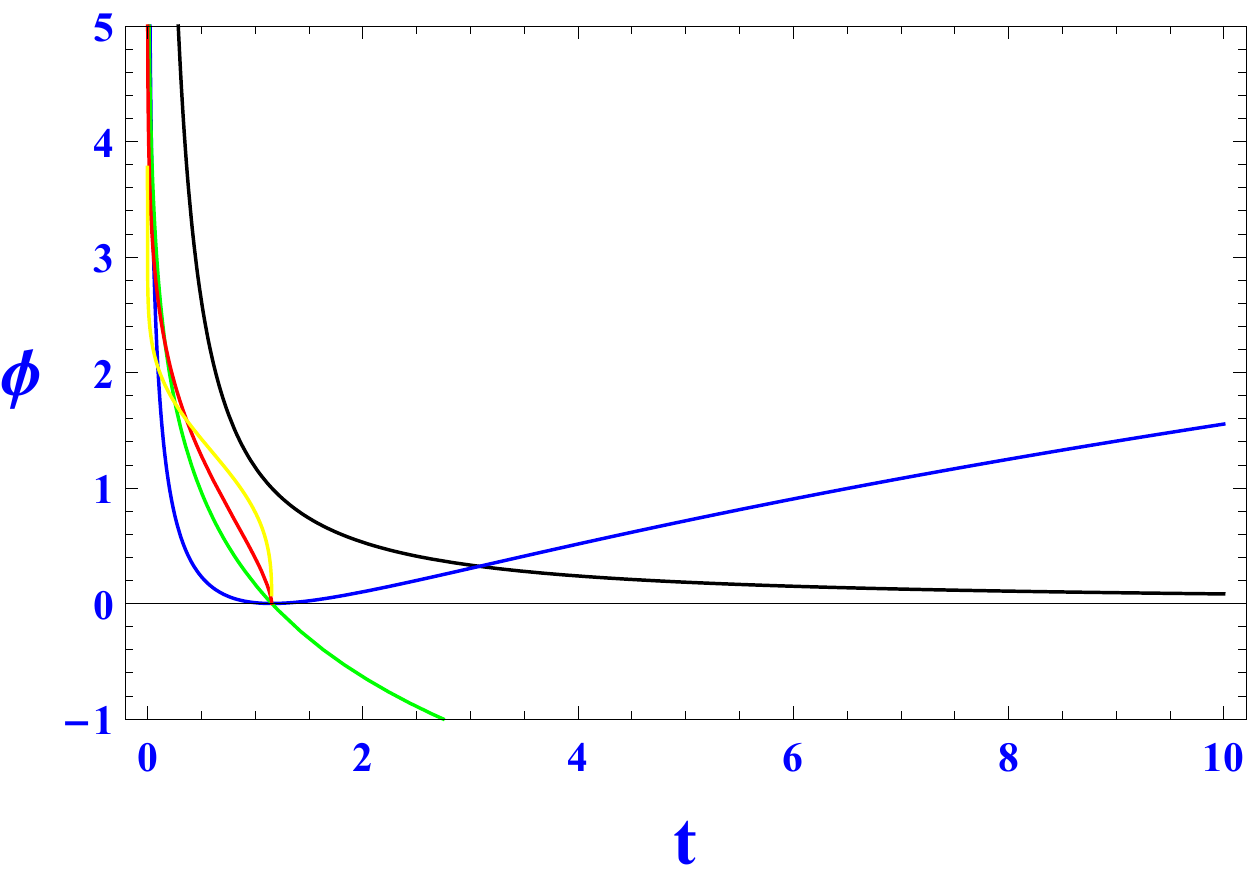}\hspace{4mm}
\caption{{\footnotesize The time behavior of the scalar field for $n=-2$
(black curve), $n=-1$ (blue curve), $n=0$ (green curve), $n=1$ (red curve) and $n=4$
(yellow curve) associated to the vacuum cosmologies.
The left and right panels correspond to the positive
and negative values of both $c_1$ and $\gamma$, respectively.
Moreover, we have taken $t_i=0$ for all cases.}}
\foreignlanguage{english}{\label{vac}}
\end{figure}
For all values of $n$, the scale factor of the universe decelerates with cosmic time.
We should note that as $a(t)\propto t^{1/3}$, disregarding the
scalar field, this solution does not correspond to
 the general relativistic spatially
flat FLRW universe in vacuum, which is Minkowski space;
whilst, it is similar to the corresponding one in
the context of BD theory, i.e., Ohanlon-Tupper
solution, when the BD coupling parameter tends to infinity.

   \subsubsection{(Effective) radiation cosmologies}
  By setting ${\rm W}_{_{\rm SB}}=\frac{1}{3}$ in (\ref{new49}) and using (\ref{new17-2}),
we get $\gamma=-\beta$ and consequently ${\cal W}=0$.
Therefore, ~(\ref{SB-5action}) reduces to the Einstein-Hilbert action for GR in five dimensions.
Consequently, as we have assumed that there is an ordinary matter in a five-dimensional space-time,
all the equations in section~\ref{Set up} reduce to those
associated to the generalized version of IMT .
We should note that, to get the correct set of solutions, we
can start from equations (\ref{ohanlon-eq-1})-(\ref{ohanlon-eq-4})
by setting $\phi={\rm constant}$.

For our model, it is straightforward to show that the
associated solutions of this effective radiation case are
\begin{eqnarray}\nonumber
 ds^2\!\!&=&\!\!dS^2|_{_{\Sigma_{o}}}=-dt^2+
 a_0^2\left(\frac{-2\beta c_1}{a_0^3 \psi_0}\right)(t-t_i)\left[dr^2+r^2\left(d\theta^2+sin^2\theta d\varphi^2\right)\right],\\\nonumber
 \psi(t)\!\!&=&\!\!\psi_0\left[\frac{-2\beta c_1}{a_0^3 \psi_0}(t-t_i)\right]^{-\frac{1}{2}},\\\nonumber
 \phi(t)&=&\phi_0={\rm constant},\hspace{10mm} V(\phi)=0,\\\nonumber
 \rho_{_{\rm SB}}\!\!&=&\!\!3p_{_{\rm SB}}(t) =\frac{3}{4\chi}(t-t_i)^{-2}\propto\frac{1}{a^4},\\
 \label{new95}
 R^{^{(4)}}\!\!&=&\!\!6(\dot{H}+2H^2)=0.
   \end{eqnarray}
Namely, the solution corresponds to the vanishing Ricci curvature.
 Moreover, we should note that the solution is independent of ${\cal W}$ and
 the extra dimension contracts as the cosmic time increases.
By comparing this particular solution obtained in our herein modified SB setting
with those obtained in the context of the standard BD theory,
we find that it is exactly the same as the Nariai solution~\cite{N63,GFR73,Faraoni.book}
associated to the radiative fluid for a spatially flat FLRW universe.

\subsubsection{(Effective) stiff fluid and false vacuum}
For the case of stiff fluid, by solving ${\rm W}_{_{\rm SB}}=1$ in (\ref{new49}) and
using (\ref{new17-2}), we find that the only acceptable solution is $\beta=0$ which
is the same value of the vacuum case already investigated.
Namely, in our model,
it is impossible to obtain a stiff fluid with
$\rho_{_{\rm SB}}=p_{_{\rm SB}}\neq0$.

For the case of false vacuum, by setting ${\rm W}_{_{\rm SB}}=-1$, we
get either $\gamma=0$ or $3\gamma+\beta=0$. The former gives the static
universe solution and it is not of interest and the latter will be investigated in the next subsection.

\subsection{Case~II:\,$3\gamma+\beta=0$}
In this case, from (\ref{new18-2}), for all values of $n$, we obtain
\begin{equation}\label{new61}
\phi^{\frac{n}{2}}\dot{\phi}=\frac{c_1}{a_0^3\psi_0}.
\end{equation}
In order to get the induced scalar potential, we substitute (\ref{new61}) into (\ref{new31}), which yields
\begin{equation}\label{new65}
\frac{dV}{d\phi}{\Biggr|}_{_{\Sigma_{o}}}=\left \{
 \begin{array}{c}
 2\beta{\cal W}\left(\frac{c_1}{a_0^3\psi_0}\right)^2\phi^{\frac{n}{2}}
 \hspace{12mm} {\rm for}\hspace{5mm} n\neq-2,\\\\
 2\beta{\cal W}\left(\frac{c_1}{a_0^3\psi_0}\right)^2\phi^{-1}
 \hspace{12mm} {\rm for}\hspace{5mm} n=-2.
 \end{array}\right.
\end{equation}
Integrating equations (\ref{new65}) gives
\begin{equation}\label{new66}
V(\phi)=\left \{
 \begin{array}{c}
-\frac{16\beta}{3(n+2)}\left(\frac{c_1\beta}{a_0^3\psi_0}\right)^2\phi^{\frac{n+2}{2}}
 \hspace{2mm} {\rm for}\hspace{5mm} n\neq-2,\\\\
 -\frac{8\beta}{3}\left(\frac{c_1\beta}{a_0^3\psi_0}\right)^2{\rm ln}\phi  \hspace{15mm} {\rm for}\hspace{5mm} n=-2,
 \end{array}\right.
\end{equation}
where we have used (\ref{Gamma-W}) and, without loss of generality, we have assumed that the
constants of integration vanish.
Consequently, by substituting each relation of (\ref{new54}) into its corresponding case in (\ref{new66}),
we get the induced scalar potential, versus cosmic time, as
\begin{equation}\label{new69}
V(t)=-\frac{8}{3}\left(\frac{c_1\beta}{a_0^3\psi_0}\right)^3\,(t-t_i),
\end{equation}
which does not depend on $n$.
Now, by employing relations (\ref{new59}), (\ref{new59-2}), (\ref{R16}), (\ref{R17})
and (\ref{new66}), the non-vanishing components of the induced matter are given by
\begin{eqnarray}\label{new74}
\rho_{_{\rm SB}}\!\!&=&\!\!\frac{1}{\chi}\left(\frac{c_1\beta}{a_0^3\psi_0}\right)^2
\left[\frac{4}{3}\left(\frac{c_1\beta}{a_0^3\psi_0}\right)(t-t_i)+1\right],\\\nonumber
\\\label{new75}
p_{_{\rm SB}}\!\!&=&\!\!\frac{1}{\chi}\left(\frac{c_1\beta}{a_0^3\psi_0}\right)^2
\left[-\frac{4}{3}\left(\frac{c_1\beta}{a_0^3\psi_0}\right)(t-t_i)+\frac{1}{3}\right],
\end{eqnarray}
where, again, we have used (\ref{Gamma-W}). By employing
(\ref{new17-2}), (\ref{new74}) and (\ref{new75}), it is easy to
show that $\dot{\rho}_{_{\rm SB}}+3H(\rho_{_{\rm SB}}+p_{_{\rm SB}})=0$.
Namely, the induced EMT for this case is also conserved.

Moreover, it is straightforward to show that
\begin{eqnarray}\nonumber
 ds^2\!\!&=&\!\!dS^2|_{_{\Sigma_{o}}}=-dt^2+
 a_0^2\,exp\left[\frac{2c_1\tilde{\beta}}{3a_0^3\psi_0}(t-t_i)\right]\left[dr^2+r^2
 \left(d\theta^2+sin^2\theta d\varphi^2\right)\right],\\\nonumber
 \psi(t)\!\!&=&\!\!\psi_0\,exp\left[-\frac{c_1\tilde{\beta}}{a_0^3\psi_0}(t-t_i)\right],\\\nonumber
\phi^{\frac{n+2}{2}}\!\!&=&\!\!\frac{c_1(n+2)}{2a_0^3\psi_0}(t-t_i)
  \hspace{21mm} {\rm for}\hspace{15mm} n\neq-2,\\\nonumber
\phi(t)\!\!&=&\!\!exp\left[\frac{c_1}{a_0^3\psi_0}(t-t_i)\right]
  \hspace{15mm} {\rm for}\hspace{15mm} n=-2,\\\nonumber
  \\\nonumber
  V(t)\!\!&=&\!\!\frac{8}{3}\left(\frac{c_1\tilde{\beta}}{a_0^3\psi_0}\right)^3\,(t-t_i),\\\nonumber
\rho_{_{\rm SB}}\!\!&=&\!\!\frac{1}{\chi}\left(\frac{c_1\tilde{\beta}}{a_0^3\psi_0}\right)^2
\left[-\frac{4}{3}\left(\frac{c_1\tilde{\beta}}{a_0^3\psi_0}\right)(t-t_i)+1\right],\\\nonumber
\\\label{new76}
p_{_{\rm SB}}\!\!&=&\!\!\frac{1}{\chi}\left(\frac{c_1\tilde{\beta}}{a_0^3\psi_0}\right)^2
\left[\frac{4}{3}\left(\frac{c_1\tilde{\beta}}{a_0^3\psi_0}\right)(t-t_i)+\frac{1}{3}\right],
\end{eqnarray}
where $\tilde{\beta}\equiv-\beta$.
The above relations indicate
 that at a fixed time as $t=t_i$ the universe commences to expand from a nonsingular value $a_0$.
 Moreover, at that fixed time, the energy density and pressure take constant values as
 $\rho_{_{\rm SB}}= 3p_{_{\rm SB}}=\frac{1}{\chi}\left(\frac{c_1\beta}{a_0^3\psi_0}\right)^2$.

 In order to describe the evolution of the universe at subsequent times, let us consider
 a few quantities by looking at the integration constants. As we are interested in the
solutions in which the fifth dimension contracts by cosmic time~\cite{OW97},
therefore, (\ref{34}) yields $c_2=-c_1\tilde{\beta}<0$.
On the other hand, let us assume that both $a_0$ and $\psi_0$ take
positive values, thus, $\frac{c_1\tilde{\beta}}{\psi_0a_0^3}$ should take positive values.

By considering relation (\ref{new74}), the above arguments
can be in accordance with $\rho_{_{\rm SB}}>0$ provided that
\begin{equation}\label{constraint}
0<\frac{4}{3}\left(\frac{c_1\tilde{\beta}}{a_0^3\psi_0}\right){\tau_{\rm f}}<1,
\end{equation}
where $\tau_{\rm f}\equiv t_{\rm f}-t_i$ is an arbitrary fixed time and
the inequality in the left hand side arises from our assumptions discussed above.
The condition (\ref{constraint}) does not give any constraint for $\frac{c_1\tilde{\beta}}{a_0^3\psi_0}$ at
early times. However, for large $\tau_{\rm f}$, it
indicates that $\frac{c_1\tilde{\beta}}{a_0^3\psi_0}<<1$. However, as this quantity is a
constant and cannot evolve with time, we conclude that it must take very small vales at all times.
We should note that these considerations have been obtained in the geometrized units where $\chi=8\pi$.

 By employing the above assumptions associated to the integration constants, we conclude
 that when the cosmic time increases, the induced energy density decreases.
 Moreover, the scale factor of the universe accelerates exponentially, which is similar to the
 one obtained in GR for the spatially flat universe, i.e., the de Sitter model.
 However, there is a crucial difference between our model and
 the de Sitter model. Namely, in our model we have $\rho_{_{\rm SB}}+p_{_{\rm SB}}=
 \frac{4}{3\chi}\left(\frac{c_1\tilde{\beta}}{a_0^3\psi_0}\right)^2\neq0$.
 Moreover, for this case, the scalar field (for $n\neq-2$) and the induced scalar potential increase when time grows.
 While $\phi(t)$ (only for $n=-2$) and $\psi(t)$ contract with the cosmic
 time, which is a sought feature in the context of induced-matter theories.



\section{Summary and Discussions}
\indent \label{conclusion}

In this paper, by employing a dimensional
reduction procedure for a five-dimensional SB theory,
we have constructed a MSBT in four dimensions.
The resulting effective EMT, namely $T_{\mu\nu}^{^{[\rm SB]}}$, contains three parts:
$T_{\mu\nu}^{^{[\rm IMT]}}$ (which is produced from the fifth dimensional part of the metric)
is geometrically induced on a four dimensional hypersurface;
$T_{\mu\nu}^{^{[\rm \phi]}}$, which has no analog
in the conventional SB theory and depends also on the scalar field and
its derivative with respect to the fifth coordinate; finally, we have an induced
scalar potential, which contributes to the SB field equations.
In the MSBT, as in the GR and the standard SB theory, the effective EMT satisfies
a prevalent conservation law. As it is seen in our herein model as well as in \cite{RM-KS17}, the modified
construction of some scalar-tensor theories provide simpler and more convenient methods for obtaining comprehensive
set of exact cosmological solutions, without taking any {\it ansatz}, onto the hypersurface,
with respect to the corresponding standard models.

It has been believed that the IMT is supported by the CM theorem. Moreover, the CM theorem has been extended
to the class of embeddings in which the embedding space-time is an Einstain space-time
rather than a Ricci-flat one, see, e.g., \cite{AL01,ADLR01,DR02-2}. In \cite{ADLR01}, instead of symmetric cases, they
have used a general treatment with full Einstein-scalar field system.
Also, they have provided general discussions concerning Cauchy as well
as initial value problems. These generalized versions of the CM theorem
might protect our herein approach in constructing the MSBT setting.
Concerning the CM theorem, two points also should be added: (i) it has been claimed
that as this theorem is only for the analytic functions, thus it
is unsuitable to employ in noncompactified theories \cite{A04}.
However, it has been
argued that as the analytic functions may only be
unsuitable to study the topological defects rather than
investigating the cosmological models where the manifold is global \cite{AB05}.
(ii) The assumption of locality in the CM
theorem was removed and a global extension of it was proved \cite{K04}.

Let us emphasize a few similarities and differences of the MSBT with respect to the
standard four dimensional SB theory. In our herein
procedure, the five-dimensional SB field equations
(\ref{(D+1)-equation-1}) and (\ref{(D+1)-equation-2}), by taking the
metric (\ref{global-metric}), produce four
sets of equations~(\ref{D2say}), (\ref{BD-Eq-DD}), (\ref{D2-phi}) and
(\ref{P-Dynamic}), such that equations~(\ref{BD-Eq-DD}) and
(\ref{D2-phi}) reproduce the standard SB field
equations, including a scalar potential on a four-dimensional
space-time, with an effective energy-momentum source.
Moreover, they can equivalently be retrieved from standard SB action including a scalar potential. Such a
correspondence is asserted by the extended versions of the CM
theorem~\cite{AL01,ADLR01,DR02-2}. However, we
should note that equation~(\ref{D2say}) has no SB analog, and the set of
equations~(\ref{P-Dynamic}) is an extended version of a conservation
law acquired in the IMT procedure.

As a case study, we have considered a five-dimensional bulk geometry and then
constructed the reduced cosmology on the four-dimensional hypersurface.
We started from a generalized FLRW line-element
and the standard SB theory in a  five-dimensional vacuum space-time.
We have shown that there are two constants of motion, which properly assisted us
to find exact solutions for the field equations.
Depending on the integration constants, we have obtained two classes of exact solutions:
in the first class (Case~I), both the scale factors $a(t)$ and $\psi(t)$ have
power-law behaviors, whereas for the second class (Case~II), they constitutes exponential functions of the cosmic time.
We should note that for both the classes, when the scale
factors are written concerning the cosmic time, they do not
depend on $n$, whilst, the scalar field does.
For $n\neq-2$, it is in the form of the
logarithmic and power-law functions of the cosmic time
for Case I and II, respectively.
While for $n=-2$, it is a power-law and exponential
functions (of cosmic time) for the Cases I and II, respectively.

Subsequently, as an application of the MSBT setting (constructed in section~\ref{Set up}), we
have used the MSBT field equations to deal with the induced (effective) quantities onto the four-dimensional hypersurface.
We should note that these quantities have interesting properties, which are presented in what follows.
We found that, for both classes of the solutions, the induced scalar potential [which is
produced from the geometry (as a fundamental concept) rather than adding it by {\it ad hoc} assumptions to the
action] as well as the components of the induced matter, are in
power-law forms (of the cosmic time) and are independent of $n$.
However, the expressions associated to the scalar field depend also on $n$ and are not always
power-law functions of the cosmic time. We have shown that the induced matter onto
the hypersurface obeys a conventional conservation law (similar
to ordinary matter in the standard SB theory and GR).

For case I, we observed that the formula for the effective matter yield a barotropic
equation of state, which is associated with a perfect fluid.
We then investigate dust, vacuum, radiation, stiff
fluid and false vacuum cosmologies. In what follows,
we would present a brief review of the results:

\begin{itemize}

   \item{\it Dust solutions:}
    For a matter-dominated universe, we have shown that
    there are two disconnected branches occur for the time ranges $t<t_i$ and $t>t_i$; such that for the
    former range, the scale factor contracts with cosmic time, while
    for the latter one, it expands. However, the extra dimension does
     behave differently. For both of the branches, we have studied
    the time behavior of the quantities with different
    values of the integration constants and $n$.

    We then focused on the branch II in which $t>t_i$.
We have shown that the scale factor of the universe decelerates with cosmic
    time, which is inapplicable for the present epoch of the universe.
    Whilst, the induced energy density, the extra dimension and the effective
    scalar potential contract with the cosmic time.
    However, to describe the time behavior of the scalar field, we
    should specify not only the integration constants but also $n$.
     We should note that for the case $n\neq-2$,
     the allowed range of the cosmic time is $t\geq t_i>0$;
     namely, our herein model does not have a big bang singularity.
     Whilst, for $n=-2$ there is a big bang singularity.
\item {\it Vacuum solutions:}
For this case, the only possibility is to set $\beta=0$, i.e., considering $\psi={\rm constant}$,
which means the components of the induced matter
and the induced scalar potential must vanish.
We showed that when $\gamma\mapsto-\tilde{\beta}/3$, features
of the scalar field associated to
this case reduce to those obtained for the dust case.
Namely, by taking suitable initial values, we get
similar time behaviors for the corresponding quantities.

We have then studied the quantities correspond to the branch II in which $t>t_i$ .
     This solution, for all values of $n$, yields a decelerating expansion for the universe.
  Despite the scale factor, the scalar field depends not
  only on the integration constants but also on $n$. We have plotted the behavior of
  the scalar field concerning cosmic time for different values of $n$ and the integration constants.
  We have shown that, for the case $n\neq-2$, there is no big bang singularity
 for the universe, but for $n=-2$, it can
 start to decelerate from a vanishing size.
  Disregarding the behavior of the scalar field and
  considering the big bang singularity, the scale factor
  behaves similar to the corresponding case in the context of the BD
  theory when the BD coupling goes to infinity.
    \item{\it Radiation solutions:} We found
    that our results correspond to the particular class of the Nariai solution
    associated to the radiative fluid for a spatially flat FLRW
    metric in the context of the BD theory. In this case, the scale factor
    decelerates, the induced scalar potential and Ricci scalar vanish,
    the scalar field takes constant values and the induced energy density, pressure
    and the extra dimension decrease with the cosmic time.

     \item{\it Stiff fluid and false vacuum:} For the case I, we found that
     the only acceptable solution for the stiff fluid is to set $\beta=0$,
     which corresponds to the vacuum case already discussed.
     Moreover, for a false vacuum, the quantity $3\gamma+\beta$ must
     vanish, which is not permitted in the case
     I, but, it corresponds to the case II.
\end{itemize}

     For the case II, we found that the scale
     factor and the components of the induced matter
take constant values at an initial time $t=t_i={\rm constant}$.
In order to investigate the behavior of the physical quantities
at subsequent times,
we have discussed the corresponding integration constants.
     Therefore, when the cosmic time increases, the scale factor of the universe accelerates
     exponentially, while the induced energy density decreases linearly such
     that it takes very small values at late times.
     These results have been obtained by assuming the geometrized units where $\chi=8\pi$.
     While increasing the cosmic time, the induced scalar potential
     increases linearly, while the behavior of the scalar field for $n=-2$
     decreases exponentially.
    Moreover, for $n\neq-2$, it is straightforward to show that
     it decreases and increases for $n<-2$ and $n>-2$, respectively.
     However, the fifth dimension always contracts with the cosmic time.
     Finally, we should note that although the solutions for case II can be
     considered as de Sitter-like, in our MSBT setting this behavior
     emerges instead from non-vanishing induced matter where $\rho_{_{\rm SB}}+p_{_{\rm SB}}=
 \frac{4}{3\chi}\left(\frac{c_1\tilde{\beta}}{a_0^3\psi_0}\right)^2\neq0$.

 Finally, we would like to mention a few points regarding our herein MSBT framework and its applications. (i) In this paper,
we have constructed the MSBT setting by going onto a four-dimensional hypersurface $\Sigma_l:l=l_0={\rm constant}$ which
is orthogonal to the fifth imension. We can generalize the embedding procedure by taking a dynamical hypersurface,
see, e.g., \cite{P06}. (ii) Instead of five-dimensional embedding space-time, we can start from an arbitrary dimensional
space-time, see, e.g., \cite{RRT95,RFM14}. Therefore, it will be possible to discuss concerning lower-dimensional gravity settings.
(iii) As the equations (\ref{D2say}) and (\ref{P-Dynamic}) lie outside the conventional SB theory, studying them may produce interesting consequences.
(iv) We have seen that the sign of the four-dimensional
Ricci curvature [see Equation (\ref{R(D+1)-R(D)})] as well as all the induced
quantities on the hypersurface depend on the signature of the five-dimensional
line-element which can also be taken as two-time metric.
We should note that applying the modified versions of the IMT-scalar-tensor settings (for instance, MBDT as well as MSBT)
 for two-time metrics may produce more interesting and generalized results than those obtained in the context of the IMT framework.


\section*{Acknowledgments}
We thank  anonymous referees for their beneficial points regarding the
embedding theorems and the two-time physics. SMMR appreciates the support of the grant SFRH/BPD/82479/2011 by
the Portuguese Agency Funda\c{c}\~ao para a Ci\^encia e
Tecnologia. This research
is supported by the grant PEst-OE/MAT/UI0212/2014.


\end{document}